\documentclass{article}
\usepackage{amssymb}
\usepackage{url}
\usepackage{a4}
\usepackage{cite}
\usepackage{theorem}
\newcommand{\hyp}{\mycal S}

\newcommand{\mcM}{{\mycal M}}

\newcommand{\bea}{\begin{eqnarray}}
\newcommand{\beaa}{\begin{eqnarray*}}
\newcommand{\bean}{\begin{eqnarray}\nonumber}
\newcommand{\bel}[1]{\begin{equation}\label{#1}}
\newcommand{\beal}[1]{\begin{eqnarray}\label{#1}}
\newcommand{\beadl}[1]{\begin{deqarr}\label{#1}}
\newcommand{\eeadl}[1]{\arrlabel{#1}\end{deqarr}}
\newcommand{\eeal}[1]{\label{#1}\end{eqnarray}}
\newcommand{\eead}[1]{\end{deqarr}}
\newcommand{\eea}{\end{eqnarray}}
\newcommand{\eeaa}{\end{eqnarray*}}

\newcommand{\be}{\begin{equation}}
\newcommand{\ee}{\end{equation}}

\newcommand{\tr}{\mbox{\rm tr}}

\newcommand{\eq}[1]{\eqref{#1}}

\newcommand{\T}{{\mathbb T}}

\DeclareFontFamily{OT1}{rsfs}{}
\DeclareFontShape{OT1}{rsfs}{m}{n}{ <-7> rsfs5 <7-10> rsfs7 <10->
rsfs10}{} \DeclareMathAlphabet{\mycal}{OT1}{rsfs}{m}{n}
\def\scri{{\mycal I}}%
\def\scrip{\scri^{+}}%
\def\scrim{\scri^{-}}%
\def\Scri{\scri}

\usepackage{amsmath} 

\def \Reel{\mathbb{R}}

\def \R {\Reel}

\def \Nat{\mathbb{N}}

\def \N {\Nat}

\newcounter{mnotecount}[section]

\newcommand{\rmnote}[1]{}

\begin{document}
\title{Recent results in mathematical relativity}
\author{
Piotr T. Chru\'sciel\thanks{Partially supported by a Polish
Research Committee grant 2 P03B 073 24, and by   the Institute of
Physics, London; email
\protect\url{Piotr.Chrusciel@lmpt.univ-tours.fr}, URL
\protect\url{ www.phys.univ-tours.fr/}$\sim$\protect\url{piotr}}\\
D\'epartement de
Math\'ematiques\\
Facult\'e des Sciences\\ Parc de Grandmont\\ F37200 Tours, France
}

\date{\today}
\maketitle
\begin{abstract}
We review selected recent results concerning the global structure
of solutions of the vacuum Einstein equations. The topics covered
include quasi-local mass,  strong cosmic censorship, non-linear
stability, new constructions of solutions of the constraint
equations,  improved understanding of asymptotic properties of the
solutions, existence of solutions with low regularity, and
construction of initial data with trapped surfaces or apparent
horizons.

This is an expanded version of a plenary lecture, sponsored by
{\em Classical and Quantum Gravity},  held at the GR17 conference
in Dublin in July 2004.
\end{abstract}

\maketitle

\section{Introduction}

Reviewing recent progress in mathematical relativity is a
difficult task, in view of the large number of excellent papers
appearing in the field. Choices have to be made because of obvious
time limits set for a lecture. In order to narrow down the number
of topics covered I will concentrate  on those results that
concern the {\em global} properties of solutions of the {\em
vacuum} Einstein equations, and which have appeared in preprint or
final form within the last three years.

\section{Quasi-local mass}

The question of localisation of mass in general relativity has a
long history, with no unanimously accepted candidate emerging so
far, see~\cite{Trautman:Witten,SzabadosLR} and references therein.
There are at least two strategies which one might adopt here:
trying to isolate a mathematically interesting object, or trying
to find a physically relevant one. In the best of the worlds the
same quantity would result, but no such thing has been found yet.
From a physical point of view the strongest case can be made, I
believe, for definitions obtained by Hamiltonian methods. Recall
that the geometric symplectic framework of Kijowski and
Tulczyjew~\cite{KijowskiTulczyjew} has been applied to general
relativity by Kijowski and
collaborators~\cite{Kijowski:variational,Kij2,KijowskiGRG,CJK,ChEinsteinKomar}),
and it allows  a systematic treatment of boundary terms, together
with associated Hamiltonians, at least at a formal
level\footnote{Kijowski's analysis leads  to symplectic structures
on spaces of fields with prescribed boundary data. To obtain a
bona fide Hamiltonian system one should prove that the resulting
initial-boundary value problems are well posed, which has not been
done so far. It would be of interest to analyse how the
Friedrich-Nagy~\cite{FriedrichNagy} initial-boundary value
problems fits into this framework.}.  One of the Hamiltonians that
emerges in this way is the following~\cite{KijowskiGRG}: Consider
a three dimensional initial data set $(M,g,K)$ in a
four-dimensional space-time $(\mcM,^4\!g)$. Let $\Sigma$ be a two
dimensional surface within $\mcM$ and suppose that the mean
extrinsic curvature vector $\kappa$ of $\Sigma $ is
\emph{spacelike}. Let {$$\lambda:=\sqrt{{}^4g(\kappa,\kappa)}$$}be
the $^4g$-length of $\kappa$. Assuming that the dominant energy
condition holds in $(\mcM,^4\!g)$, it follows from the embedding
equations that the Gauss curvature of the metric induced by $g$ on
$\Sigma$ is positive. One can then invoke the Weyl embedding
theorem~\cite{Nirenberg:Weyl,Pogorelov} to isometrically embed
$(\Sigma,^4\!g|_{\Sigma})$ into $\R^3$. We shall denote by
$\lambda_0$ the associated $\lambda$
 as
calculated using the flat metric in $\R^3\subset \R^{3,1}$. Let
$m_{\scriptsize \mathrm{K}}$ be the Kijowski mass of $\Sigma$,
\bel{Kijowski}\displaystyle m_{\scriptsize \mathrm{K}}= \frac 1 {8\pi}
\int_{\Sigma} (\lambda_0-\lambda)d^2\mu\;.\ee
 A surprising theorem of Liu and Yau~\cite{LiuYau} asserts that
 $$m_{\scriptsize \mathrm{K}}\ge0 \;,$$ \emph{with equality if and only
if $(M,g,K)$ is a subset of Euclidean $\R^3\subset \R^{3,1}$}. The
key to the proof is
 a similar result by Shi and Tam~\cite{ShiTam}, which is the Riemannian
analogue  of this statement: Shi and Tam prove that for manifolds
of positive scalar curvature, the mean curvature $H$ of a convex
surface bounding a compact set satisfies
\bel{BY}\displaystyle m_{\scriptsize \mathrm{BY}}= \frac 1 {8\pi}
\int_{\Sigma} (H_0-H)d^2\mu\ge 0\;.\ee Here $H_0$ is the mean
curvature of an isometric embedding of $\partial M$ into $\R^3$
(thus $H_0$ coincides with $\lambda_0$). Liu and Yau show that the
inequality \eq{Kijowski} can be reduced to the Shi-Tam inequality
using Jang's equation, in a way somewhat similar to the transition
from the ``Riemannian" to the ``full" Schoen-Yau positive mass
theorems~\cite{schoen:yau:ADM,SchoenYau81}. The ``quasi-local
mass" $m_{\scriptsize \mathrm{BY}}$ appearing in \eq{BY} has been
introduced, and studied, by Brown and
York~\cite{YorkBrown,LauYorkBrown}.

In~\cite{OMST} O'Murchadha, Szabados and Tod show that
$m_{\scriptsize \mathrm{K}}\ne 0$ for some surfaces in Minkowski
space-time, which indicates that the normalisation in
\eq{Kijowski} is not optimal; this leaves room for future
improvements.

The celebrated papers by Bray~\cite{Bray:preparation2} and Huisken
and Ilmanen~\cite{HI2} proving the  Riemannian Penrose inequality
have appeared recently  (see also~\cite{ChBray} and references
therein). Those papers settle the problem for initial data sets
with \emph{positive Ricci scalar} (this is a restrictive condition
which is satisfied, e.g., for maximal initial data sets), proving
that the ADM mass is not less than the square root of the area of
the outermost minimal surface divided by  $4\pi$. Suggestions
how to approach the \emph{full} Penrose inequality, without the
$R\ge 0$ restriction, have recently been made by Frauendiener and
by Malec, Mars and Simon in~\cite{MMS,Frauendiener:Penrose}.
In~\cite{MalecOM:Jang} Malec and \'O Murchadha show that a direct
approach based on Yang's equation cannot succeed.
In~\cite{WeinsteinYamada} Weinstein and Yamada point out that  for
\emph{connected} charged black holes, with charge $Q$ and with
$R=\sqrt{A/4\pi}$ -- the area radius of the outermost minimal
surface, a Penrose-type inequality involving global charge
\bel{WY} m\ge \frac 12 \left(R + \frac {Q^2} R \right)\ee follows
from the Huisken-Ilmanen proof. They also prove that \eq{WY} fails
for some initial data sets with two black holes; the proof is yet
another application of the Corvino-Schoen perturbation technique,
discussed in Section~\ref{SCS} below.

We finish this section by noting some recent improvements in our
understanding of the mass of asymptotically hyperbolic
manifolds~\cite{Wang,ChNagyATMP,CJL,ChHerzlich}. Recall, now, that
the asymptotically flat positive mass theorem can be used to prove
uniqueness of static  asymptotically flat black
holes~\cite{bunting:masood}; similarly the positive mass theorem
for asymptotically hyperbolic manifolds can be used to prove
uniqueness results in this context~\cite{AndDahl,BGH}
(compare~\cite{ACD,ACD2,GSW,GSW2} for completely different
approaches). A particularly elegant proof of rigidity of
hyperbolic space, reducing the problem to the standard
asymptotically Euclidean positive energy theorem, has been
recently given by Qing~\cite{Qing:uniqueness}. The approach  has
been exploited in~\cite{BMQ} by Bonini, Miao and Qing to
considerably extend the rigidity result. Yet another recent
approach to rigidity of asymptotically hyperbolic manifolds, based
on volume comparison, has been presented by Shi and Tian
in~\cite{ShiTian}.

\section{Strong cosmic censorship}

The \emph{strong cosmic censorship (SCC) problem} concerns
{predictability}:  it is a fundamental requirement that solutions
of good physical theories should be \emph{uniquely determined by
initial data}. This is not the case for Einstein equations --- in
general relativity there exist examples where uniqueness fails
\cite{Misner,ChImaxTaubNUT}. In this context, a key result is a
theorem of Choquet-Bruhat and Geroch~\cite{ChoquetBruhatGeroch69},
which states that \emph{to any initial data one can associate,
uniquely up to a diffeomorphism, a maximal globally hyperbolic
development of those data}. The resulting space-time is unique in
the class of globally hyperbolic space-times, but in some
situations it can be extended in more than one way to strictly
larger vacuum solutions. In such cases the extension always takes
places across a null hypersurface called \emph{Cauchy horizon.}

 A mathematical formulation of strong cosmic censorship, essentially due
to Moncrief and Eardley \cite{EM} ({\em cf.\/} also
\cite{SCC,ChrCM,RendallLiving}), is the following:
\begin{quote}
{\em Consider the collection of initial data for, say, vacuum or
electro--vacuum space--times, with the initial data surface $\hyp$
being compact, or with the initial data
\newcommand{\umac}{\gamma}%
\newcommand{\metrict}{{}^3g }%
$(\hyp,\metrict,K)$ 
--- asymptotically flat. For generic such data
the maximal globally hyperbolic development thereof is inextendible.}
\end{quote}

The reader is referred to~\cite{Andrev} for an excellent recent
review of SCC.

Because of the difficulty of the strong cosmic censorship problem,
a full understanding of the issues which arise in this context
seems to be completely out of reach at this stage. There is
therefore some interest in trying to understand that question
under various restrictive hypotheses, {\em e.g.}, symmetry. Such a
program has been initiated by Moncrief in
\cite{Moncrief:Gowdy,EM}, and some further results in the
spatially compact case have been obtained in
\cite{ChIM,isenberg90,SCC,ChRendall}. The simplest case, of
spatially homogeneous space-times, has turned out to be
surprisingly difficult, because of the intricacies of the dynamics
of some of the Bianchi models. Spectacular progress in the
understanding of this last class of solutions has been made a few
years ago by Ringstr\"om. His results imply, amongst others, that
curvature blows up in the contracting direction in
 generic Bianchi models. This forbids Cauchy horizons (this was already known, by completely different
 abstract arguments, from the work
 in~\cite{ChRendall}), and also provides further information about
 the geometry near ``the end of space-time".

The next simplest case if that of Gowdy metrics on
$\T^3:=S^1\times S^1 \times S^1$: in coordinates $t\in
(-\infty,0)$ and $(\theta,x^1,x^2)\in \T^3$ the metrics are of the
form
\beaa g&=& e^{-\gamma/2}\mid t\mid ^{-1/2}(-dt^2+d\theta^2) + \mid t\mid e^P
(dx^1)^2+ 2 \mid t \mid e^PQ \,dx^1dx^2
\\ && +
\mid t \mid \left(e^PQ^2+e^{-P}\right)(dx^2)^2\;, \eeaa
with
 $\partial_{x^1}$ and $\partial_{x^2}$ being  Killing
vectors.
 The essential part of the evolution equations consists of two coupled non-linear equations:
\beaa &\displaystyle
\partial_t^2 P -
\partial^2_\theta P = - \frac {\partial_t P }{t} + e^{2P}
\left((\partial_t Q)^2 -(\partial_\theta  Q)^2\right) \;, & \\
&\displaystyle \partial_t^2 Q - \partial^2_\theta Q = - \frac
{\partial_t Q }{t} -2 \left(\partial_t P \partial_t Q
-\partial_\theta P\partial_\theta Q\right) \;. & \eeaa The
question of SCC in this class of metrics has been settled by
Ringstr\"om~\cite{RingstroemMiami}, who proved that the set of
smooth initial data for Gowdy models on $\T^3$ that do \emph{not}
lead to the formation of Cauchy horizons contains a set which is
open and dense within the set of all smooth initial data. Some key
steps in Ringstr\"om's analysis are provided by the analysis of
Fuchsian PDEs of Kichenassamy and
Rendall~\cite{KichenassamyRendall,Rendall:2000ih}, and the
analysis of the action of Geroch transformations by Rendall and
Weaver~\cite{RendallWeaver} (compare~\cite{ChCh2}). See
also~\cite{ChLake} for the related problem of an exhaustive
description of Cauchy horizons in those models.

Now, the existence of two Killing vectors is also compatible with
$S^3$, $L(p,q)$ (``lens" spaces), and $S^1\times S^2$ topologies.
Thus, to achieve a complete understanding of the set of spatially
compact initial data with precisely two Killing vectors one needs
to extend Ringstr\"oms analysis to those cases. There is an
additional difficulty that arises because of the occurrence of
axes of symmetry, where the ($1+1$)--reduced equations have the
usual singularity associated with polar coordinates. Nevertheless,
in view of the analysis by Christodoulou and
Tahvildar-Zadeh~\cite{CT93,CT293} (see also~\cite{ChANoP}), the
global geometry of \emph{generic} maximal globally hyperbolic
solutions with those topologies is reasonably well understood.
This leads one to expect that one should be able to achieve a
proof of SCC in those models using simple abstract arguments, but
this remains to be seen.

Recall, finally, that general models with two Killing vectors
$X_1$ and $X_2$  on $\T^3$ have non-vanishing \emph{twist
constants}:
$$c_a=\epsilon_{\alpha\beta\gamma\delta} X_1^\alpha X_2^\beta \nabla^\gamma X_a^\delta\;, \qquad a=1,2\;.$$
The Gowdy metrics are actually ``zero measure" in the set of all
$U(1)\times U(1)$ symmetric metrics on $\T^3$ because $c_a\equiv
0$ for the Gowdy models. The equations for the resulting metrics
are considerably more complicated when the $c_a$'s do not vanish,
and only scant rigorous information is available on the global
properties of the associated solutions~\cite{BCIM,IW}. It seems
urgent to study the dynamics of those models, as they are expected
to display~\cite{BIW} highly oscillatory behavior as the
singularity is approached. Thus, they should provide the simplest
non-isotropic model in which to study this behavior.

There has also been progress in the understanding of models with
\emph{exactly one} Killing vector. Here one only has
\emph{stability} results in the expanding direction, away from the
singularity: In~\cite{ChBCargese} Choquet-Bruhat considers  $U(1)$
symmetric initial data for the vacuum Einstein equations on a
manifold of the form $\Sigma\times S^1$, where $\Sigma$ is a
compact surface of {genus} ${ g>1}$. One assumes that the initial
data have constant mean curvature and are sufficiently close to
$(g_0,K_0)$, where $g_0$ is a product metric
$$g_0=h+dx^2\;,$$ with $h$ --- a metric of constant Gauss
curvature on $\Sigma$, and with $K_0$ --- pure trace. The sign of
the trace of $K_0$ determines an expanding time direction and a
contracting one. Under those conditions, Choquet-Bruhat proves
that the solution exists for an {infinite proper time} in the
expanding direction. The analysis builds upon previous work by
Choquet-Bruhat and Moncrief~\cite{01713418}, where a supplementary
polarisation condition has been imposed. Nothing is known in the
{contracting} direction in those models, where mixmaster behavior
is expected~\cite{belinskii71b,BGIMW}.

The proof of the above result bears some similarity to the
\emph{future stability} theorem of {Andersson and
Moncrief}~\cite{AndMon} for a class of hyperbolic models
\emph{without any symmetries}. Those authors consider initial data
near  a negatively curved compact space form satisfying a specific
rigidity condition, with the extrinsic curvature being close to a
multiple of the metric,  obtaining future geodesic completeness in
the expanding direction. The control of the solution is obtained
by studying the Bel-Robinson tensor, and its higher-derivatives
analogues. One of the  ingredients of the proof is the use of an
elliptic-hyperbolic system of equations to obtain local existence
in time~\cite{AnderssonMoncriefAIHP}.

This last reference is interesting in its own right because of the
following:  the standard local existence theory for the general
relativistic Cauchy problem proceeds through two steps. First one
solves the equations in some chosen gauge, e.g.\ in harmonic
coordinates. As a second step one patches the resulting solutions
together to construct the space-time manifold. When working with
harmonic coordinates, this second step leads to an increase by one
of the differentiability threshold for existence and uniqueness of
the solution. The elliptic-hyperbolic system
of~\cite{AnderssonMoncriefAIHP} is expected to solve that problem,
the elliptic character of some of the equations providing improved
regularity of the solution. There is, however, some more work
needed to settle this issue, related to the fact that Andersson
and Moncrief use a constant-mean curvature (CMC) slicing. The
current state-of-the art uniqueness
theory~\cite{bartnik:variational} for CMC slices requires, roughly
speaking, $C^{1,1}$ metrics, and this provides a threshold for the
analysis involved. It is conceivable that some weak form of the
maximum principle such as, e.g., that in~\cite{AGHCPAM}, could be
used to lower the $C^{1,1}$ threshold, but this remains to be
seen.

\section{Stability of Minkowski space-time}

One of the flagship results in mathematical general relativity is
nonlinear stability of Minkowski space-time, first proved by
{Christodoulou and Klainerman}~\cite{Christodoulou:Klainerman} on
some 500 pages of their celebrated book. An alternative approach
to the problem, due to {Klainerman and
 Nicol\`o}~\cite{KlainermanNicoloBook} has appeared a year ago in print. That proof is based on an analysis of
outgoing and ingoing null cones, and takes around 380 pages. A
completely new argument of {Lindblad and
Rodnianski}~\cite{LindbladRodnianski} is available now. Their
proof takes some  65 Latex pages, though it should be said that
the conditions on the initial data there are much stronger than in
previous work: those authors require that the initial data
coincide with Schwarzschildian ones outside of a compact set. We
note that under that asymptotic conditions, together with a
smallness condition, the global existence follows already from the
stability results of Friedrich~\cite{Friedrich}. However, it has
been announced by the authors (private communication) that the
method is flexible enough to allow the inclusion of a scalar
field, and to handle the following, rather weak, asymptotic
behavior of the initial data:
\bel{dec} g=(1+2M/r)\delta + O(r^{-1-\alpha})\;,\qquad  K=
O(r^{-2-\alpha})\;.\ee It should be emphasised that the decay
conditions \eq{dec} are much weaker than those
in~\cite{Christodoulou:Klainerman,KlainermanNicoloBook}.

\section{Initial data engineering}
\label{SCS}

General relativistic initial data sets have to satisfy the
constraint equations, \bel{consm} \left(
\begin{array}{c}
J\\
  \\
\rho
\end{array}
\right) (K,g):= \left(
\begin{array}{l}
2(-\nabla^jK_{ij}+\nabla_i\;\tr  K)\\
  \\
R(g)-K^{ij}K_{ij} + (\tr  K)^2
\end{array}
\right) = \left(
\begin{array}{l}
0\\
  \\
0
\end{array}
\right)\;. \ee This makes it difficult to construct space-times
with controlled global properties, such as geodesically complete
space-times with conformally smooth asymptotics (Penrose's
``asymptotic simplicity"), or solutions containing
 many black holes, or wormholes, or trapped
surfaces, or apparent horizons, and so on. A new technique for
deforming initial data sets has been invented by Corvino and
Schoen~\cite{CorvinoSchoenprep,Corvino}. One of the applications
of that technique is the unexpected
statement~\cite{CorvinoSchoenprep} that
\begin{quote}
\emph{asymptotically flat initial data with well defined mass,
momentum, centre of mass, and angular momentum can be deformed in
the asymptotic region to a Kerr metric, with arbitrarily small
change in global Poincar\'e charges}
\end{quote}
 (compare~\cite{ChDelay}).
 While this theorem is very interesting in its own, what is even more important is the
 new technique introduced, which has already led to a few noteworthy applications.
One of them is \emph{existence of asymptotically simple
space-times}: recall that a space-time is \emph{asymptotically
simple}\/~\cite{penrose:scri} if it has smooth conformal structure
at null infinity $\Scri$, and if all maximally extended null
geodesics have initial and final end points on $\scri$.  Examples
of such space-times include Minkowski space-time, and static
asymptotically flat stars, and the static
 solutions of the  Einstein -- Yang-Mills equations of Bartnik and
 McKinnon~\cite{BartnikMckinnon88}.
More generally, spaces-times which are stationary and vacuum
outside of a world tube have a smooth conformal structure at null
infinity~\cite{Dain:2001kn,Damour:schmidt}, but the
Kruskal-Szekeres-Schwarzschild space-time is not asymptotically
simple  because of null geodesics terminating in the singularities
under the event horizons. In any case, asymptotically simple
space-times possess a smooth global $\Scri$, and various previous
attempts to construct asymptotically simple \emph{vacuum}
solutions have been unsuccessful. For instance, the $C$-metrics
have both singularities inside space-time and at
$\Scri$~\cite{AshtekarDray,BicakSchmidt}; the Robinson-Trautman
metrics possess a smooth $\Scri^+$ which is complete to the
future, but which is expected to be complete to the past only for
the Schwarzschild metric; the space-times constructed by
Christodoulou-Klainerman are not known to possess smooth
asymptotic structure. In fact, the only previous dynamical example
satisfying reasonable field equations
 has been given by  Cutler and Wald, in electro-vacuum~\cite{CutlerWald}.

The paper~\cite{ChDelay2} gives the first proof of \emph{existence
of non-trivial asymptotically simple vacuum  space-times.} The
construction uses the Corvino-Schoen gluing
techniques~\cite{ChDelay} and Friedrich's stability
theorem~\cite{Friedrich}, and  the simplest version of the result
is the following: One considers a non-trivial  initial data set
 $(\R^3,\mathring g,\mathring K=0)$, with ADM mass $\mathring m$,
 satisfying a \emph{parity condition}
{\bel{parity}
\mathring g_{ij}(x)= \mathring g_{ij}(-x)\;.\ee}  One also assume
that one is given on  $\R^3\setminus B(R)$ a parity-symmetric
 {\em reference family}  $(\R^3, g^m,K^m=0)$: this is, by definition, a family
of metrics labeled by their ADM masses $m$, such that the  $m$'s cover continuously a
neighborhood of $\mathring m$. In~\cite{ChDelay2} it is proved
that \begin{quote}
\emph{the initial data set $(\R^3,\mathring g,0)$ can be
deformed to {\em some} member of the reference family on
$B(2R)\setminus B(R)$ if the metrics are close enough near
$S(R)$}.
\end{quote}If
   {$\mathring m$ is sufficiently small one can invoke Friedrich's stability theorem
   to obtain existence of a global smooth
$\Scri$, as well as asymptotic simplicity}.

We note that large families of parity-symmetric, scalar flat
metrics $\mathring g$, with mass $\mathring m$ as small as
desired, can be constructed using the conformal
method~\cite{ChDelay}. Similarly, the collection of  reference
families that can be used here is rather large: one
  can  use Schwarzschild metrics, or Weyl metrics,
or other static metrics~\cite{Reula:static,Klenk}. In fact, any
metric on $\R^3$ with zero scalar curvature and with ADM mass $m$
equal to $ \mathring m$ is part of a reference family obtained by
scaling the metric and the coordinates, $$x^i \to \lambda
x^i\;,\quad g\to \lambda^{-2} g\;\qquad\Longrightarrow \qquad m\to
\lambda^{-1} \mathring m\;.$$ The stationary character of the
reference metrics is needed to be able to assert smoothness of
$\scri$, but is not needed in the gluing construction.

The time-symmetry condition $K=0$ has only been made for
simplicity of presentation of the result, and the construction
also provides families of asymptotically simple space-times which
are not time-symmetric.

The parity condition \eq{parity} does look rather unnatural at
first sight. It can be replaced by the requirement that
$\mathring{g}$ is sufficiently close to a metric which satisfies
\eq{parity}. It is related to the fact that in the gluing process
one needs to adjust the centres of mass of the metrics which are
being glued. Parity guarantees that the {centres of mass of all
metrics involved are automatically zero}, and therefore there is
nothing to adjust.  There is a simple {Newtonian analogue which
illustrates the problem:}  Let $\rho$ be the Newtonian energy
density, $m$ the Newtonian mass (integral over the ``space
manifold" $M=\R^3$ of $\rho$), and let $\vec c$ be the Newtonian
centre of mass of the gravitating system,
$$\vec c = \int _M  \vec x \rho \;.$$
 If $\vec x$ is changed to $\vec x + \vec a$, then the new
centre of mass $\vec c\,'$ equals
$$\vec c\,' = \int _M  (\vec x + \vec a)\rho = \vec c + m \vec a \;.$$
In order to obtain $\vec c\, '=0$, one needs to translate the
system by {$-\frac{\vec c}m$}, which can be very large even if
$\vec c$ is small when $m$  is small. Because the gluing
techniques are  based on the implicit function theorem, they break
down when the translations cease to be very small.

So far we have been gluing metrics which were close to each other,
with the final metric not being drastically different from the
original one. It turns out that  gluings can lead to objects with
global properties rather distinct from the ones we started with.
The idea here is to make a small gluing to add a ``neck" to a
first initial data set $(M_1,g_1,K_1)$, and then another small
gluing to add a second initial data $(M_2,g_2,K_2)$ set on the
other end of the neck. In this way one obtains a \emph{new initial
data set on the connected sum manifold $M_1\# M_2$}. A similar
construction adds wormholes within a single initial data set. This
has been successfully carried out by Isenberg, Mazzeo and
Pollack~\cite{IMP,IMP2} a few years ago, using the conformal
method. Their approach works for CMC initial data (and, with an
additional hypothesis, for data which is only CMC near the neck
region) in either the compact, asymptotically flat or
asymptotically hyperbolic setting. The conformal factor can be
made as close to one as desired away from the necks by making the
necks small enough, but those techniques lead to   deformations of
the original data sets which are never localised. The question
then arises, whether this can be done by deformations which are
supported in a small neighborhood of the neck.

It is easily seen that there is an obstruction do to that, related
to the positive energy theorem: if one could glue, via local
deformations, a non-trivial initial data set to $(\R^n,\delta,0)$,
where $\delta$ denotes the Euclidean metric, one would end up with
a non-trivial initial data set with zero ADM mass, which is
impossible.

Nevertheless, in~\cite{CIP} it is shown that
\begin{quote}
\emph{generic}
 initial data sets can be glued together, with \emph{no deformations
 away from arbitrarily small sets} where the gluing takes place.
\end{quote}
Here a data set is   \emph{generic} if there are no  \emph{local
Killing vectors} in the associated space-time. (A local Killing
vector field is  a Killing vector field defined only on a subset
of the space-time, not necessarily globally. The intuitively
obvious fact, that non-existence of local Killing vectors is a
generic condition, is rigorously justified in~\cite{ChBeignokids}
in several important cases.) The proofs use a version with
boundary of the results in~\cite{IMP,IMP2}, the results by
Bartnik~\cite{bartnik:variational} on existence of maximal
surfaces, together with Corvino-Schoen type
techniques~\cite{CorvinoSchoenprep,ChDelay}.

One interesting application of this gluing technique is the
construction of spatially compact, \emph{vacuum}, maximal globally
hyperbolic space-times without any CMC surfaces~\cite{CIP,CIPPRL}.
The argument follows an idea of
Bartnik~\cite{bartnik:cosmological}, who obtains such space-times
which contain dust.

The Corvino-Schoen gluing technique is based on an analysis of the
operator which arises by linearising the map $(g,K)\mapsto
(J,\rho)$ of \eq{consm}, and it is well known that the properties
of this map determine the manifold structure of the set of
solutions of the constraint equations~\cite{FischerMarsden79}. The
analysis in that last reference allows one to introduce a
Fr\'echet manifold structure on solutions of the constraint
equations away from initial data sets with symmetries, but fails
to provide a Banach manifold structure. In a recent beautiful
paper~\cite{bartnik:phase} Bartnik manages to overcome the
technical difficulties that arise, obtaining a Hilbert manifold.
Various different Banach manifold structures have also been
recently constructed in~\cite{ChDelayHilbert}, again as an
application of the Corvino-Schoen techniques; further applications
of the technique can be found in~\cite{WeinsteinYamada,ChDelay}.

\section{Asymptotic structure}

The following {questions have been keeping mathematical
relativists busy for years:}
\begin{quote}{\em What are the asymptotic conditions on the initial data {at
$\{t=0$\}} that guarantee that the resulting space-time has
\begin{enumerate}
\item``a piece of smooth $\scrip$"? \item a global smooth
$\scrip$?
\end{enumerate}}
\end{quote}
As already pointed out,  Corvino-Schoen gluings lead to initial
data with a maximal globally hyperbolic development containing a
{smooth piece of $\scri$} when the resulting initial data are
 {stationary} at large distances (and even global smooth $\scri$'s for small
initial data). They provide thus the first example of non-trivial
initial data which are relevant in this context. The
Christodoulou-Klainerman theorem~\cite{Christodoulou:Klainerman}
gives a $\scri$ with very low differentiability, not enough for
 {peeling theorems}. It is conceivable that one shouldn't care
 about that, the properties of $\scri$ in Christodoulou-Klainerman space-times being enough for
e.g. for the Trautman-Bondi mass-loss formula. However, for the
sake of completeness of our understanding of the asymptotics of
the gravitational field, one would like to have an answer to
questions \emph{1} and \emph{2}.

Some progress on this has been recently made by Klainerman and
Nicol\`o~\cite{KlainermanNicoloPeeling}, who show that
\emph{peeling holds for initial data with asymptotics}
\bel{KNdec}{ g=(1+\frac{M}{2r})^4\delta + O(r^{-3-\alpha})\;,\qquad
K= O(r^{-4-\alpha})\;,\quad \alpha>0\;.}\ee Large classes of
initial data as above, not necessarily stationary outside of a
compact set, can be constructed by a variant of the Corvino-Schoen
gluing~\cite{ChDelay2}. We emphasise that \eq{KNdec} provides
\emph{peeling}, but does not guarantee smoothness of $\scri$. In
fact, the \emph{first known explicit obstructions to smoothness of
$\scri$ in terms of data at $t=0$} have been found by
Valiente-Kroon~\cite{Valiente-Kroon:2002fa,ValienteKroon:2003ux,ValienteKroon:2003ix,ValienteKroon:2004pu}.
This leads to the striking statement, that
\begin{quote}
\emph{Bowen-York} and \emph{Brill-Lindquist} initial data do {not} lead
to a smooth $\scri$ (if any).
\end{quote}

In Friedrich's Carg\`ese lecture
notes~\cite{FriedrichCargese,FriedrichFanfreluche} and
in~\cite{Valiente-Kroon:2002fa} the following conjecture is made:
There exists $N(k)\in \N$, with $N(k)\to\infty$ if $k\to\infty$,
such that
\begin{quote}
\emph{time-symmetric} initial data lead to a $\scri$ of
differentiability ${C^k}$ if and only if the data are {\em static}
up to terms $O(r^{-N(k)})$.
\end{quote}
One expects an analogous statement without the time-symmetry
condition, but this requires more care. As discussed
in~\cite{ValienteKroon:2004pu},  smoothness of $\scrip$ imposes
conditions on the initial data that are not identical to those for
smoothness of $\scrim$. The requirement that \emph{both} $\scrip$
and $\scrim$ are $C^k$ is expected then to be compatible only with
those initial data which are \emph{stationary} up to terms
$O(r^{-N(k)})$, for some $N(k)\to_{k\to\infty}\infty$.

So far we have been assuming that the cosmological constant
vanishes. Some new results  by Rendall on existence and properties
of conformal completions for vacuum space-times with $\Lambda\ge
0$ can be found in~\cite{Rendall:2003ks,Rendall:2004kh}. Recall
that Friedrich's analysis~\cite{Friedrich} applies only to
dimension four, while Rendall's approach works in higher
dimensions as well. A new method of approaching the question of
existence of conformal completions in \emph{all even space-time
dimensions} has been presented by Anderson in~\cite{AndersonCIE}.

 \section{Low regularity solutions}

Standard theory of hyperbolic
equations~\cite{HughesKatoMarsden77,Taylor96} shows that local
existence of solutions of the general relativistic Cauchy problem
holds for initial data  {$(g,K)\in H_{s}\times H_{s-1}$ with
$s>5/2$}. There are several reasons why it is of interest to try
to lower this threshold. An improved existence theorem:
\begin{itemize}
\item renders certain singularities innocuous
\item can potentially be used to simplify and/or extend global existence results
\item gives insight into the mathematical structure of the equations.
\end{itemize}
A first step in this direction has been taken by Klainerman and
Rodnianski~\cite{Klainerman:2001gt,Klainerman:2001gu}, who show
that the existence time of the solution can be bounded from below
by a function that depends only upon
$$\mathbf\|(g,K)\|_{H_{2+\epsilon}\times H_{1+\epsilon}}$$
 where $\epsilon$ is as small as desired. (See~\cite{BahouriCheminBeijing,TataruBeijing} and references
therein for related results  by Bahouri and Chemin, and by Tataru,
and others, for large classes of nonlinear wave equations.)

The existence of initial data in the
 {${H_{2}\times H_{1}}$} class is proved by
 {Choquet-Bruhat in~\cite{Choquet-Bruhat:safari}  and by
 {Maxwell in~\cite{Maxwell:2004yb}}. (In fact, Maxwell's paper goes further and produces
initial data in $H_{s} X H_{s-1}$ for $s> 3/2$, and shows that one
can approximate (in that topology) this data by smooth data.) The
already mentioned construction of manifold
 structure by Bartnik~\cite{Bartnik} allows ${H_{2}\times H_{1}}$
 differentiability as well. Klainerman conjectured that
\begin{quote}
\emph{$L^2$ regularity of the Riemann tensor should suffice for
existence.}
\end{quote}
Some partial results towards the proof of this conjecture can be
found in the paper~\cite{Klainerman:2003wx} by
 Klainerman and Rodnianski.

\section{Vacuum initial data set with apparent horizons and trapped
surfaces}

A classical statement in Hawking and Ellis~\cite{HE} asserts that
\emph{the boundary of a trapped region is an apparent horizon.}
The argument given there does not quite prove that, as it seems to
make the implicit hypothesis that the boundary of the trapped
region is a $C^2$ surface, and it is a still an open question
whether or not this is correct. Now, the main point  is the
existence of an apparent horizon within $\hyp$, and this  has been
shown to be correct very recently by Schoen~\cite{SchoenMiami}:
\begin{quote}\emph{Suppose that an asymptotically flat initial data set $(\hyp,g,K)$ contains
a compact trapped surface  $\mycal T$, then there exists in $\hyp$
a smooth, outermost, compact apparent horizon enclosing $\mycal
T$.}
\end{quote} The proof proceeds by constructing
singular solutions of the Jang equation, with delicate control of
their blow-up set. This generalises a previous result by Huisken
and Ilmanen~\cite{HI2}, who prove the corresponding statement in
the time-symmetric case.

We finish this section, and this review, by noting that Dain
\cite{Dain:ah}  and {Maxwell~\cite{Maxwell:ah}} have provided
direct constructions of initial data with smooth, trapped or
marginally trapped, boundaries.

\medskip

\noindent {\sc Acknowledgements} \  Useful comments by D.~Pollack
on a previous version of this manuscript are gratefully
acknowledged.


\def\cprime{$'$}
\providecommand{\bysame}{\leavevmode\hbox
to3em{\hrulefill}\thinspace}
\providecommand{\MR}{\relax\ifhmode\unskip\space\fi MR }
\providecommand{\MRhref}[2]{%
  \href{http://www.ams.org/mathscinet-getitem?mr=#1}{#2}
} \providecommand{\href}[2]{#2}

\end{document}